\documentclass[%
 aip,
 jmp,%
 amsmath,amssymb,floatfix,
 reprint,%
]{revtex4-1}

\usepackage[version=3]{mhchem} %

\usepackage{dcolumn}%
\usepackage{bm}%
\usepackage{todonotes}
\usepackage{tikz}
\usepackage{nicefrac}

\newcommand{\vk}[1]{{\color{black}{#1}}}
\newcommand{\mc}[1]{{\color{black}{#1}}}
\newcommand{\braket}[1]{\left<{#1}\right>}

\begin{document}

\title{High Order Path Integrals Made Easy
}%

\author{Venkat Kapil}
 \affiliation{Laboratory of Computational Science and Modelling, Institute of Materials, Ecole Polytechnique F\'ed\'erale de Lausanne, Lausanne, Switzerland}
 
\author{J\"org Behler}
\affiliation{Lehrstuhl f\"ur Theoretische Chemie, Ruhr-Universit\"at Bochum, Bochum, Germany}

\author{Michele Ceriotti}
 \email{michele.ceriotti@epfl.ch}
\affiliation{Laboratory of Computational Science and Modelling, Institute of Materials, Ecole Polytechnique F\'ed\'erale de Lausanne, Lausanne, Switzerland}%

\date{\today}%

\begin{abstract}

The precise description of quantum nuclear
fluctuations in atomistic
modelling is possible by employing path integral techniques,
which involve a considerable computational overhead due to the 
need of simulating multiple replicas of the system. 
Many approaches have been suggested
to reduce the required number of replicas. 
Among these, high-order factorizations
of the Boltzmann operator are particularly attractive
for high-precision and low-temperature scenarios. 
Unfortunately, to date several technical challenges have prevented
a widespread use of these approaches to study 
nuclear quantum effects in condensed-phase systems. 
Here we introduce an inexpensive molecular dynamics scheme 
that overcomes these limitations, thus making it 
possible to exploit the improved convergence of 
high-order path integrals without having to sacrifice
the stability, convenience and flexibility of conventional
second-order techniques.
The capabilities of the method are demonstrated 
by simulations of liquid
water and ice, as described by a neural-network potential fitted
to dispersion-corrected hybrid density 
functional theory calculations.

\end{abstract}

\maketitle

\section{Introduction}

Molecules and materials that contain light nuclei - most notably hydrogen - 
exhibit considerable deviations from classical behavior, which are most pronounced
at cryogenic temperatures but extend up to and even above room temperature~\cite{raug-klei03jacs,mora+13prl,ceri+16cr}.
Examples of such nuclear quantum effects (NQEs) 
include a heat capacity that deviates
dramatically from the Dulong-Petit limit~\cite{deby12adp}, equilibrium fractionation
of isotopomers between different phases of a given compound~\cite{bern+00sci} or different molecular
sites~\cite{webb+16gca}, non-Maxwell-Boltzmann
distribution of particle velocities~\cite{andr+05advp} as well as dynamical
properties that differ from the predictions obtained from classical
molecular mechanics~\cite{habe+13arpc}. 

The interest in modelling NQEs in atomistic simulations has 
been growing constantly over the past years -- not only because
faster computers and more efficient algorithms have made such
tasks more accessible, but also because of a paradigm change 
regarding the employed inter-atomic potentials.
In particular, ab initio simulations 
solving explicitly the electronic structure problem
at increasingly accurate levels of theory ``on-the-fly''~\cite{delb+12jctc} as well as next-generation 
potentials~\cite{behl-parr07prl,bart+10prl,medd+14jctc}, that are designed to reproduce ab initio reference data
rather than experiments, have made the need of modelling NQEs
an urgent matter. This is because in the absence of empirical fitting parameters,
which is a mandatory condition for predictive simulations,
these ab initio studies employ the Born-Oppenheimer
energy surface, without any term that could implicitly compensate for the lack
of zero-point energy and tunnelling. Therefore, for predictive simulations it is 
insufficient to focus on the inter-atomic forces alone.
NQEs also need to be included
explicitly on a sound physical basis to achieve the highest possible accuracy. 

The techniques of choice for treating the quantum nature of nuclei are
path integral molecular dynamics (PIMD), and path integral Monte Carlo (PIMC) simulations~\cite{cepe95rmp}.
Using an elegant mapping of the quantum mechanical partition function onto
the classical partition function of an extended system composed of several
replicas of the atomic configuration, path integral methods make it possible
to treat exactly the quantum statistics of distinguishable~\cite{parr-rahm84jcp} (and indistinguishable~\cite{chan-woly81jcp,cepe95rmp})
particles, however at a much larger cost compared to a classical simulation. 

In the past few years, several approaches have been proposed to reduce the overhead
of such simulations by accelerating the convergence with the number of replicas. 
This goal has been achieved by computing expensive parts of the potential
on a reduced number of replicas~\cite{mark-mano08jcp,mark-mano08cpl}, \vk{by using a thermostat described by a generalized Langevin equation (GLE)} to artificially generate the proper quantum fluctuations~\cite{ceri+09prl2,ceri+11jcp,ceri-mano12prl}, or by using a higher-order
expansion of the quantum partition function -- so-called ``high-order path integral
techniques''~\cite{taka-imad84jpsj,suzu95pla,chin97pla,shinichi}. This latter approach is very appealing, particularly if low-temperature
or high-accuracy are sought, 
since the convergence improves with the number of replicas $P$ from $P^{-2}$ of traditional methods to $P^{-4}$.
Unfortunately, the higher order expansion introduces some cumbersome terms in the forces 
that depend formally on the Hessian of the physical potential. Therefore, rather than integrating
the high-order PIMD equations of motion directly, research has focused on re-weighting
schemes~\cite{jang-voth01jcp,yama05jcp,mars+14jctc}, which are however affected by
statistical inefficiency that worsens as system size increases~\cite{ceri+12prsa}.
A truncated cumulant expansion has recently shown considerable promise, although
\emph{ad hoc} estimators need to be devised specifically for different system properties~\cite{polt-tkat16cs}.

In the present work we will demonstrate that performing full high-order PIMD can be achieved using a symplectic finite-difference integrator. In this approach there is only a modest computational overhead, that is quickly paid off when high-accuracy or low-temperature simulations are to be performed. Moreover, we will show that it is possible to combine high-order path integrals with colored-noise acceleration techniques, although in practice
there is only a small advantage relative to GLE techniques applied on top of second-order PIMD. We will demonstrate the capabilities of the method using liquid water described by a neural network (NN) potential~\cite{behl-parr07prl,morawietz2016} as an example.

\section{Methods}

\subsection{Second-order and fourth-order path integrals}

The most effective framework for treating the quantum
mechanical behavior of distinguishable particles is 
based on the path integral formalism, that maps
the quantum mechanical partition function
$Z = {\rm tr}[e^{-\beta \hat{H}}]$ at
 the inverse temperature $\beta=\frac{1}{k_{\rm B} T}$, 
onto a classical partition
function in an extended ``ring polymer'' phase space.
This mapping corresponds to the application of
the identity $e^{-\beta \hat{H}}=\left[e^{-\beta \hat{H}/P}\right]^P$,
followed by a high-temperature expansion of the 
Boltzmann operator $e^{-\beta_P \hat{H}}$ -- where we have introduced
the shorthand notation $\beta_P=\beta/P$. The most 
commonly used approach relies on a Trotter decomposition in terms of the potential operator
$\hat{V}$ and the kinetic energy operator $\hat{T}$,
\begin{equation}
e^{-\beta \hat{H}} \approx  [e^{-\beta_P \frac{\hat{V}}{2}} e^{-\beta_P \hat{T}}e^{-\beta_P \frac{\hat{V}}{2}}]^{P},
\label{eq:tr_split}
\end{equation}
which leads to an expansion of the (low temperature)
Boltzmann operator that is accurate up to second
order in $\beta_P$. 
It is easy to show that the application of the Trotter splitting
yields a classical-like partition function.
For a system of $N$ distinguishable particles with masses $\{m_i\}$ evolving under a potential $V(\mathbf{q}_{1},\dots,\mathbf{q}_{P})$ at an inverse temperature of $\beta$, 
the ring polymer Hamiltonian reads $\mathcal{H}_{P}^{\text{tr}}(\mathbf{p},\mathbf{q}) =
\mathcal{H}_{P}^{0}(\mathbf{p},\mathbf{q}) + \mathcal{V}_P(\mathbf{q})$. Here, the free ring polymer Hamiltonian is 
\begin{equation}
   \mathcal{H}_P^0(\mathbf{p},\mathbf{q}) = \sum_{i=0}^{N-1} \sum_{j=0}^{P-1} \left(\frac{[\mathbf{p}_i^{(j)}]^2}{2m_i} + \frac{1}{2} m_i \omega_P^2 [\mathbf{q}_i^{(j)} - \mathbf{q}_i^{(j+1)}]^2\right)
\label{eq:free-rp}
\end{equation}
and the physical potential term $\mathcal{V}_P(\mathbf{q})$ 
is just a sum
over the potential $V$ evaluated for the various replicas
\begin{equation}
    \mathcal{V}_P(\mathbf{q})= \sum_{j=0}^{P-1} V(\mathbf{q}_1^{(j)},...,\mathbf{q}_N^{(j)}).
\label{eq:total-pot}
\end{equation}
The ``beads'' in the ring polymer are connected cyclically 
(i.e. $j+P \equiv j$ in Eq.~\eqref{eq:free-rp}) by springs of 
frequency $\omega_P = 1/\beta_P$. 
Sampling
has to be performed at the inverse temperature $\beta_P$. 

High-order path integral schemes rely on a more accurate
decomposition of the density matrix than in 
Eq.~\eqref{eq:tr_split}, which however requires 
including terms that depend on the commutator
between $\hat{V}$ and $\hat{T}$. Several of these
schemes have been proposed~\cite{taka-imad84jpsj}, 
all however 
having similar advantages and shortcomings.
We will focus in particular on the Suzuki-Chin (SC)
decomposition~\cite{suzu95pla,chin97pla}, that is accurate
up to fourth order in $\beta_P$, and reads
\begin{equation}
    e^{-\beta \hat{H}} \approx  [e^{-\beta_P \frac{\hat{V_e}}{3}} e^{-\beta_P \hat{T}} e^{-\beta_P \frac{4\hat{V_o}}{3}} e^{-\beta_P \hat{T}} e^{-\beta_P \frac{\hat{V_e}}{3}}]^{\frac{P}{2}},
\label{eq:sc_split}
\end{equation}
where
\begin{align}
    & \hat{V}_e = \hat{V} + \frac{\alpha}{6} \beta_P^2 [\hat{V},[\hat{T},\hat{V}]], \\
    & \hat{V}_o = \hat{V} + \frac{(1-\alpha)}{12} \beta_P^2 [\hat{V},[\hat{T},\hat{V}]].
\end{align}
$\alpha \in [0,1]$ is an arbitrary parameter 
and can be adjusted to improve the convergence for 
a given problem. It seems, however, that no 
generally-applicable prescription for its choice
can be obtained. In the present study, for 
reasons that will become apparent later on, 
we always used $\alpha=0$. The main advantage
of the SC scheme is that any structural observable
can be computed seamlessly by averaging over the 
even beads in the path, without the complex correction
terms that often enter estimators in other high-order schemes. 

Following the same procedure as in the Trotter
case, one can obtain a classical partition function
based on the splitting~\eqref{eq:sc_split}, 
which is accurate up to fourth order in $\beta_P$. 
The Suzuki-Chin ring-polymer Hamiltonian
$\mathcal{H}_P^\text{sc} \left(\mathbf{p},\mathbf{q}\right) = \mathcal{H}_P^\text{0} + \mathcal{V}^\text{sc}_P(\mathbf{q})$ contains a modified
potential term that acts differently on odd and even
beads,
\begin{equation}
   \mathcal{V}^\text{sc}_P(\mathbf{q})=
   \sum_{j=0}^{P-1}
     \left(w_j V\left(\mathbf{q}^{(j)}\right) + \sum_{i=0}^{N-1}\frac{w_j d_j}{m_i \omega_P^2}\left|\mathbf{f}_i^{(j)}\right|^2\right),
     \label{eq:sc_v}
\end{equation}
where $\mathbf{f}_i^{(j)}=-\partial V\left(\mathbf{q}^{(j)}\right)/\partial\mathbf{q}_i^{(j)} $ is the physical 
force acting on the $i$-th atom in the $j$-th replica, and the 
scaling factors for odd and even beads are given by
\begin{equation}
\begin{alignedat}{2}
& w_j=2/3,\quad d_j={\alpha}/{6} \quad && \text{$j$ is even},  \\
& w_j=4/3, \quad d_j={\left(1-\alpha\right)}/{12} \quad && \text{$j$ is odd}. 
\end{alignedat}%
\label{eq:sc-wd}
\end{equation}
Estimators for  the Suzuki-Chin propagator can be derived by two 
routes. ``Thermodynamic'' (TD) estimators are obtained by applying thermodynamics identities to the SC ring polymer partition function, 
while ``operator'' (OP) estimators result from carrying out the splitting
operation on an expression that already contains the quantum mechanical operator. 
The latter class of estimators is generally simpler to derive
and evaluate, but in the
present manuscript we will compare potential and kinetic
energy operators computed using both routes.
Expressions
for both TD and OP-method estimators have been derived and
reported several times~\cite{jang-voth01jcp,pere-tuck11jcp}, 
but we also list them in Appendix~\ref{sec:estimators} for
the sake of completeness.

Evaluating the modified potential~\eqref{eq:sc_v} 
only requires knowledge of the first derivative of the
physical potential, which in a PIMD simulation has
to be computed to evolve the dynamics. 
However, evaluating the force associated with 
$\mathcal{V}^\text{sc}_P$ is not so trivial,
as it contains second derivatives of the 
potential,
\begin{equation}
\begin{split}
    \mathbf{f}_{i}^{\text{sc},(j)}  
    &= w_j (\mathbf{f}_{i}^{(j)} + \frac{2 d_j}{\omega_P^2} \tilde{\mathbf{f}}_{i}^{(j)}) =\\
    &=
    w_j \mathbf{f}_{i}^{(j)} + \frac{2 w_j d_j}{\omega_P^2} \sum_{k=0}^{N-1} \frac{\partial^2 V(\mathbf{q}^{(j)})}{\partial{\mathbf{q}_i^{(j)}} \partial{\mathbf{q}_k^{(j)}} }\frac{\mathbf{f}_{k}^{(j)}}{m_k}.
\end{split}
\end{equation}

\subsection{Finite-differences Suzuki-Chin PIMD}

One should notice, however, that the 
expression for $\tilde{\mathbf{f}}$ involves
the second derivative of $V$ 
\emph{projected on the mass-scaled force}. 
As it has been recognized in the context
of high-order path integral Monte 
Carlo~\cite{pred04pre,buch-vani13cpl}, and similarly
to what has been done for instance in 
electronic structure theory~\cite{putr+00jcp}, it
is possible to evaluate this kind of 
projected second derivatives by finite 
differences (FD), \mc{using either a symmetric estimator
\begin{equation}
\tilde{\mathbf{f}}_{i}^{(j)}=
\lim_{\epsilon\rightarrow 0}
\frac{1}{2\epsilon\delta} \left[
{\mathbf{f}}_{i}^{(j)}\left(\mathbf{q}^{(j)}+
\epsilon\delta\, {\mathbf{u}}^{(j)}\right)-
{\mathbf{f}}_{i}^{(j)}\left(\mathbf{q}^{(j)}-
\epsilon\delta\, {\mathbf{u}}^{(j)}\right)
\right]
\label{eq:finitediff}
\end{equation}
or an asymmetric FD formula
\begin{equation}
\tilde{\mathbf{f}}_{i}^{(j)}=
\lim_{\epsilon\rightarrow 0}
\frac{1}{\epsilon\delta} \left[
{\mathbf{f}}_{i}^{(j)}\left(\mathbf{q}^{(j)}+
\epsilon\delta\, {\mathbf{u}}^{(j)}\right)-
{\mathbf{f}}_{i}^{(j)}\left(\mathbf{q}^{(j)}\right)
\right],
\label{eq:fw-finitediff}
\end{equation}
}
where 
\begin{equation}
{\mathbf{u}}^{(j)}_i=\frac{{\mathbf{f}}^{(j)}_i}{m_i},\quad \delta=\left[\frac{1}{NP}\sum_k \frac{ {\mathbf{f}}^{(j)}_k\cdot {\mathbf{f}}^{(j)}_k}{m_k^2} \right]^{-1/2}
\label{eq:fd-displacement}
\end{equation}
indicate a displacement that is parallel to the mass-scaled force,
and a normalization coefficient so that 
$\epsilon$ corresponds to the root mean
square displacement applied on each atom when 
computing the derivative. 

The expression in Eq.~\eqref{eq:finitediff} can
be used seamlessly to propagate the equations of
motion and to evaluate the estimators
for thermodynamic and structural properties listed in 
Appendix~\ref{sec:estimators}.
The crucial aspect that makes this procedure
viable is that using \mc{any of the two
finite-difference estimators} for the derivative
yields a rigorously
time-reversible and symplectic integrator, when combined with 
a symmetric Trotter split
velocity-Verlet integrator (see Appendix~\ref{sec:sympl}). 
As a result, the scheme is stable even for relatively
large values of the finite-difference step, which
is advantageous e.g. when evaluating the forces in \emph{ab initio} calculations, where residual errors in the convergence
of the self-consistent solution to the electronic
structure problem inevitably lead to noisy forces. 
It is also important to note that, when setting $\alpha=0$, 
${\tilde{\mathbf{f}}^{(j)}}$ only needs to be evaluated for
odd beads, so that in practice the forward-backward 
evaluation of the derivatives costs as much as
one full force evaluation, making this approach
twice as expensive as a Trotter PIMD simulation
with the same number of beads.  \mc{ As we will 
show later, the asymmetric finite-difference
estimator appears to be only marginally less stable
than its symmetric counterpart, which means that 
with a judicious choice of $\epsilon$ one can 
reduce the number of force 
evaluations  by 50\%{}.\footnote{In this work we could
afford parallelizing calculations over all of the
beads. As a consequence, there would be no 
advantage in leaving half of the processors idle
during the evaluation of ${\tilde{\mathbf{f}}^{(j)}}$.
Therefore, we used throughout the symmetric FD
expression.}  }
This relatively small overhead can be reduced even
further by using a multiple-time step  (MTS) integrator 
for $\tilde{\mathbf{f}}$ (see Appendix~\ref{sec:mts-sc}),
and is quickly compensated
by the much faster asymptotic convergence. 
Our finite-difference Suzuki-Chin scheme is already
advantageous at room temperature, and its lead 
becomes substantial for low-temperature or 
high-accuracy studies. 
Similar expressions can be easily derived in the 
context of other high-order factorizations such as
that introduced by Takahashi and 
Imada~\cite{taka-imad84jpsj}, and 
it is possible that perturbed path 
estimators~\cite{polt-tkat16cs} could be derived
on top of a full fourth-order path integral Hamiltonian,
providing even faster convergence to quantum
expectation values. The availability of projected
force derivatives also facilitates the
implementation of the fourth-order version of estimators for
the heat capacity~\cite{yama05jcp} and for
isotope fractionation ratios~\cite{chen-ceri14jcp}. 
Finally, further dramatic speed-ups can be obtained
whenever one  can apply range-separation 
techniques such as 
ring-polymer contraction~\cite{mark-mano08jcp},
since we have made sure that our implementation 
in i-PI~\cite{ceri+14cpc} is fully compatible with 
that of conventional real and imaginary-time multiple 
time stepping~\cite{kapi+16jcp}.

\subsection{A generalized Langevin equation for
high-order path integrals}

Having access to direct sampling of $\mathcal{H}_P^\text{sc}$
 opens up the possibility of combining high-order
path integrals with a generalized Langevin equation 
acceleration. In the Trotter case, the
\vk{normals mode (NM)} eigenvectors of the Hamiltonian
for a harmonic potential
$V(q)=m\omega^2q^2/2$ do not depend on the 
frequency $\omega$ itself, that only leads to a shift
to the \vk{NM} frequencies. This makes 
it possible to apply sophisticated thermostatting
strategies, with different GLEs applied to 
individual \vk{NM} coordinates -- all without the 
need of knowing the \vk{NM} decomposition of the physical potential~\cite{ceri-mano12prl}.
Unfortunately, this is not the case for 
the Suzuki-Chin Hamiltonian. 
However, since the NM transformation
remains an orthogonal transformation,
it is possible to apply a single GLE
to all Cartesian (or Trotter NM) 
coordinates which gives the same
effect as applying such GLE onto
the proper SC \vk{NMs}.
Starting from this observation, 
one can -- with considerable effort, 
see Appendix~\ref{sec:scgle} --
derive a frequency-dependent effective
temperature $T^\star(\omega)$, that enforces
different fluctuations on different 
ring-polymer vibrations so as to 
obtain converged quantum expectation values
for \emph{any} OP-method estimator of 
position-dependent properties in the 
harmonic limit, and for any number of 
beads. Contrary to the Trotter case, where one
can further tune ring-polymer fluctuations
to accelerate the convergence of the 
centroid-virial kinetic energy estimator,
this is not possible here, so we can expect that 
the convergence of the quantum kinetic energy will be
less efficient than with the PIGLET approach~\cite{ceri-mano12prl}.
GLE parameters enforcing the desired temperature curve
for this Suzuki-Chin GLE (SC+GLE) approach have been
obtained following the fitting protocol discussed in Ref.~\cite{ceri+10jctc}, 
and are available for download from an on-line repository~\cite{GLE4MD}.

\section{Neural Network Water: a Benchmark}

For a comprehensive benchmark study of the methods discussed in the previous section
we will use simulations of water, \vk{a} prototypical system for the 
investigation of nuclear quantum effects. For this purpose we use a neural network (NN)
potential fitted to ab initio calculations performed with the B3LYP
hybrid density functional~\cite{beck93jcp} and the D3 dispersion corrections by Grimme~\cite{grim+10jcp}, 
as implemented in CP2K~\cite{vand+05cpc}. \vk{The potential is fully reactive, i.e. it allows for the possibility of bond breaking and formation, and } 
has recently been shown to provide an
excellent description of nuclear quantum effects in water, as probed by 
isotope fractionation and the nuclear quantum kinetic energy~\cite{chen+16jpcl}, at the same time allowing  us to obtain thorough sampling. The potential was evaluated
using a  NN implementation~\cite{RunnerLAMMPS} for 
LAMMPS~\cite{plim95jcp}. 
Unless otherwise specified, 
each result we report involved a trajectory of at least 200~ps for a supercell containing 128 molecules at the experimental density. We enforced constant-temperature sampling at $T=300$~K using a PILE-G scheme~\cite{ceri+10jcp}
with  $\gamma_k=\omega_k/2$, and a weak, global thermostat on the centroid  -- so that effectively canonical-sampling runs correspond to the thermostatted ring-polymer molecular dynamics (TRPMD)  protocol ~\cite{ross+14jcp} suitable to discuss quantum dynamical properties. In order to probe the behavior of our approach in a lower-temperature regime, we also performed simulations of a 96-molecules box of hexagonal ice at $T=100$~K. For colored-noise simulations we used the PIGLET thermostat~\cite{ceri-mano12prl} for Trotter PI, and the SC+GLE strategy discussed above for SC PIMD. These calculations will be a challenging test case for our techniques, because the reactive nature of the NN potential allows for quantum fluctuations of the hydrogen bond probing the strongly anharmonic regions in the potential energy surface of water~\cite{ceri+13pnas}. 

\begin{figure}[hbtp]
\includegraphics[width=1.0\columnwidth]{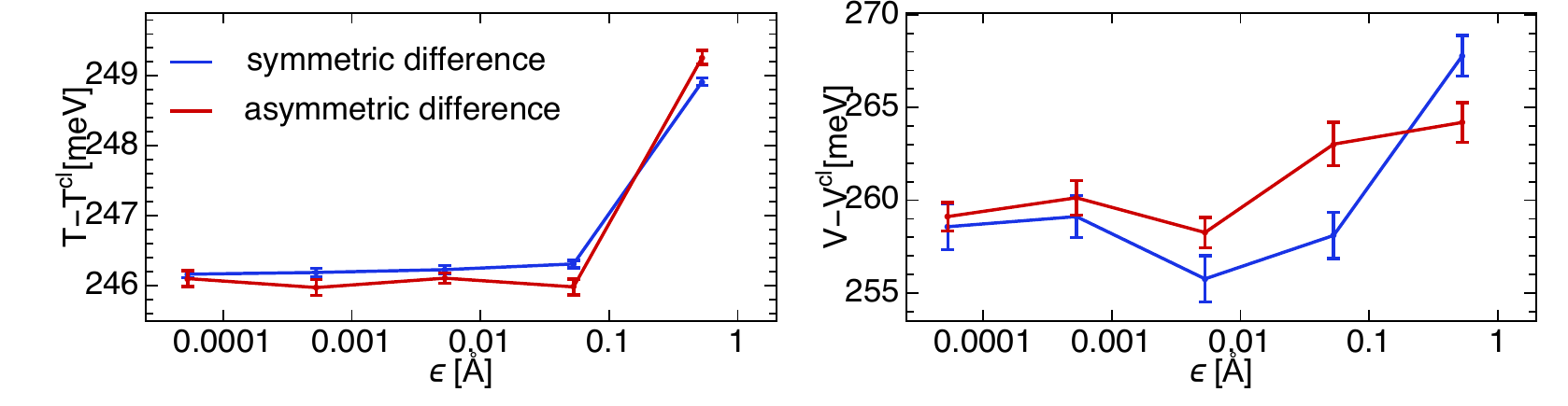}
\caption{\label{fig:stability}
Expectation values of the quantum contributions
\vk{per molecule} to the potential energy $V$ as well as to the 
kinetic energy $T$, as a function
of the finite-difference displacement $\epsilon$,
for a SC-PIMD simulation of liquid water at $300$~K performed with 16 beads. 
\mc{The two sets of points correspond to 
the symmetric (blue) and the asymmetric (red) 
finite-difference integrators.}
Error bars indicate the statistical error, which is 
of the order of 1\%{} for the potential and of the 
order of 0.1\%{} for the kinetic energy. 
The superscript ``cl'' indicates the classical component.
Note that given the definition of the displacement 
vector $\mathbf{u}$ in Eq.~\eqref{eq:fd-displacement}, the finite-difference step $\epsilon$ indicates the 
root-mean-squared displacement of an atom
during the evaluation of the derivative. 
}
\end{figure}

\subsection{Stability of the finite-difference scheme}

A possible problem that one has to be aware of when using force evaluation schemes based on discrete approximations of the derivatives is that in many cases
-- most notably for \emph{ab initio} simulations -- imperfect convergence of 
self-consistency schemes can introduce numerical noise. 
In particular, when using a small displacement
in a finite difference scheme, the signal-to-noise ratio degrades,
which can lead to instabilities in the 
integration of the equations of motion. 
Therefore it is important to
test how sensitive are the results to the specific 
value of the atomic displacement. 
Figure~\ref{fig:stability} shows that the \vk{SC integrators
we introduce here, due to their time-reversibility 
and symplectic properties, show} exceptionally good 
stability, with no appreciable effect of the 
root mean squared atomic displacement  
on the quantum expectation
values of the potential and kinetic energies
for $\epsilon\lesssim 0.1$\AA{}.
\mc{Over this broad range of 
displacements there is no appreciable difference
between the symmetric and the forward FD estimator,
so that the latter should be used whenever one does
not parallelize fully the calculation over the $P$ 
beads.}

\begin{figure}[hbtp]
\includegraphics[width=1.0\columnwidth]{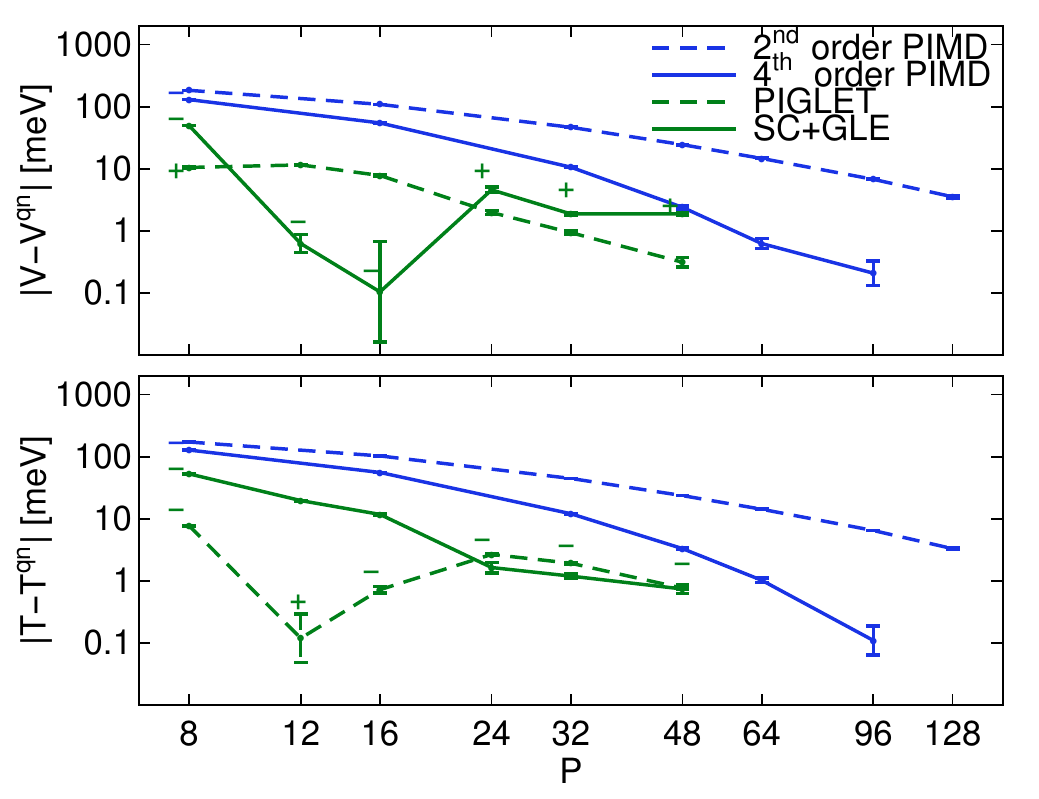}
\caption{\label{fig:energies}
Error \vk{per molecule} on the value of the potential energy $V$ as well as the 
kinetic energy $T$, as a function
of the number of beads $P$, for a simulation of 
liquid water at $300$~K performed with second and fourth-order PIMD, and with the 
corresponding colored-noise methods, namely PIGLET (second order) and SC+GLE (fourth order). 
We only report here the OP-method estimators -- 
 see Figure~\ref{fig:tdop} for a comparison with the TD estimators.
 The fully-converged value is taken to be
 SC PIMD with $P=48$, and errors are plotted on a log-log scale to highlight the faster convergence of fourth-order methods.
}
\end{figure}

\subsection{Convergence of energy estimators}

\begin{figure}[hbtp]
\includegraphics[width=1.0\columnwidth]{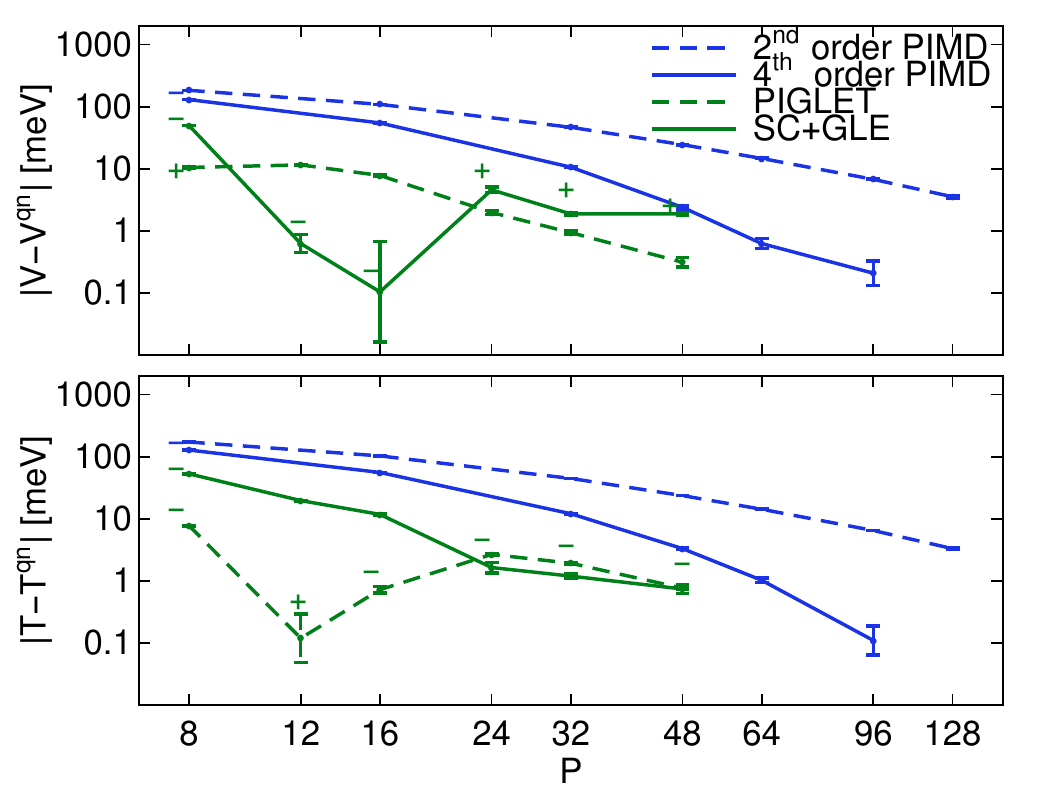}
\caption{\label{fig:ice}
Error \vk{per molecule} on the value of the potential energy $V$ as well as the 
kinetic energy $T$, as a function
of the number of beads $P$, for a simulation of 
ice at $100$~K performed with second and fourth-order PIMD, and with the 
corresponding colored-noise methods, namely PIGLET (second order) and SC+GLE (fourth order). 
We only report here the OP-method estimators -- 
 see Figure~\ref{fig:tdop} for a comparison with the TD estimators.
 The fully-converged value is taken to be
 SC PIMD with $P=128$, and errors are plotted on a log-log scale to highlight the faster convergence of fourth-order methods.
}
\end{figure}

The most straightforward measure for the convergence of a PIMD
method to the quantum limit is given by the potential and 
kinetic energy estimators. 
Figure~\ref{fig:energies} shows such convergence tests, comparing
Trotter and SC path integrals with
and without colored noise. Results are
in line with the expectations. Fourth-order 
PIMD gives a much improved asymptotic convergence, 
without the statistical instabilities 
observed in re-weighting strategies~\cite{ceri+12prsa}
and giving with $P=16$ results that are
superior to Trotter PI with $P=32$. 
\mc{The number of evaluation of 
$\tilde{\mathbf{f}}_{i}^{(j)}$ can be reduced
with a MTS scheme  (see Appendix~\ref{sec:mts-sc}).
Even by computing the SC force 
as often as every $M=2$ steps,}
it can be clearly seen
that also at room temperature our
finite-differences implementation
of high-order path integrals
provides higher accuracy at a 
smaller cost than Trotter PIMD.
As shown in figure~\ref{fig:ice},
the improvement becomes 
even more significant as the 
temperature is lowered. In 
a simulation of hexagonal ice at
$T=100K$, SC PIMD reaches
an error of a few meV per molecule when $P=48$.
When using Trotter PIMD, one would need to use
more than 128 beads to obtain a similar accuracy.

GLEs improve significantly the convergence 
of both standard and fourth-order PIMD,
giving potential energies that
are within a few percent of the converged
results with as few as 4-6 beads
for water, and 16-24 beads for 
ice. Although the \vk{GLE-thermostatted}
results are better than the 
\vk{canonically sampled} PI simulations for
all values of $P$, we observe that
the convergence of
GLE techniques is non-monotonic, 
similar to what was observed in 
simulations of small molecules
at ultra-low temperature~\cite{uhl+16jcp}. 
It appears that the convergence of
SC+GLE is not better than that 
obtained by PIGLET, which 
underscores the fact that the 
limiting factor for convergence
of GLE schemes has more to do
with zero-point energy leakage
between different modes than
with the asymptotic convergence
of the PI section of the method.
SC+GLE results are more sensitive to 
the coupling strength of the colored 
noise than in the case of PIGLET,
probably due to the more complex
form of the full path integral
Hessian in the harmonic limit
(see Appendix~\ref{sec:scgle}).

Although the possibility of combining 
high-order path integrals
with correlated noise sampling
might be beneficial in some 
\vk{ specific cases -- for instance when 
computing structural properties
at ultra-low temperature --} it seems that
the best course of action should
be to use PIGLET whenever an
accuracy of a few percent is 
sufficient, and resort to SC PIMD
with conventional thermostatting
whenever one wants \vk{(a)} to  reach the
ultimate level of convergence,
\vk{(b)} to use 
sampling techniques (e.g. replica
exchange) for which it is necessary
to have a well-defined functional
form for the phase-space density,
or \vk{(c)} to compute 
\vk{complicated} estimators whose convergence is  not
accelerated by GLEs. 

\begin{figure*}[hbtp]
\includegraphics[width=1.0\textwidth]{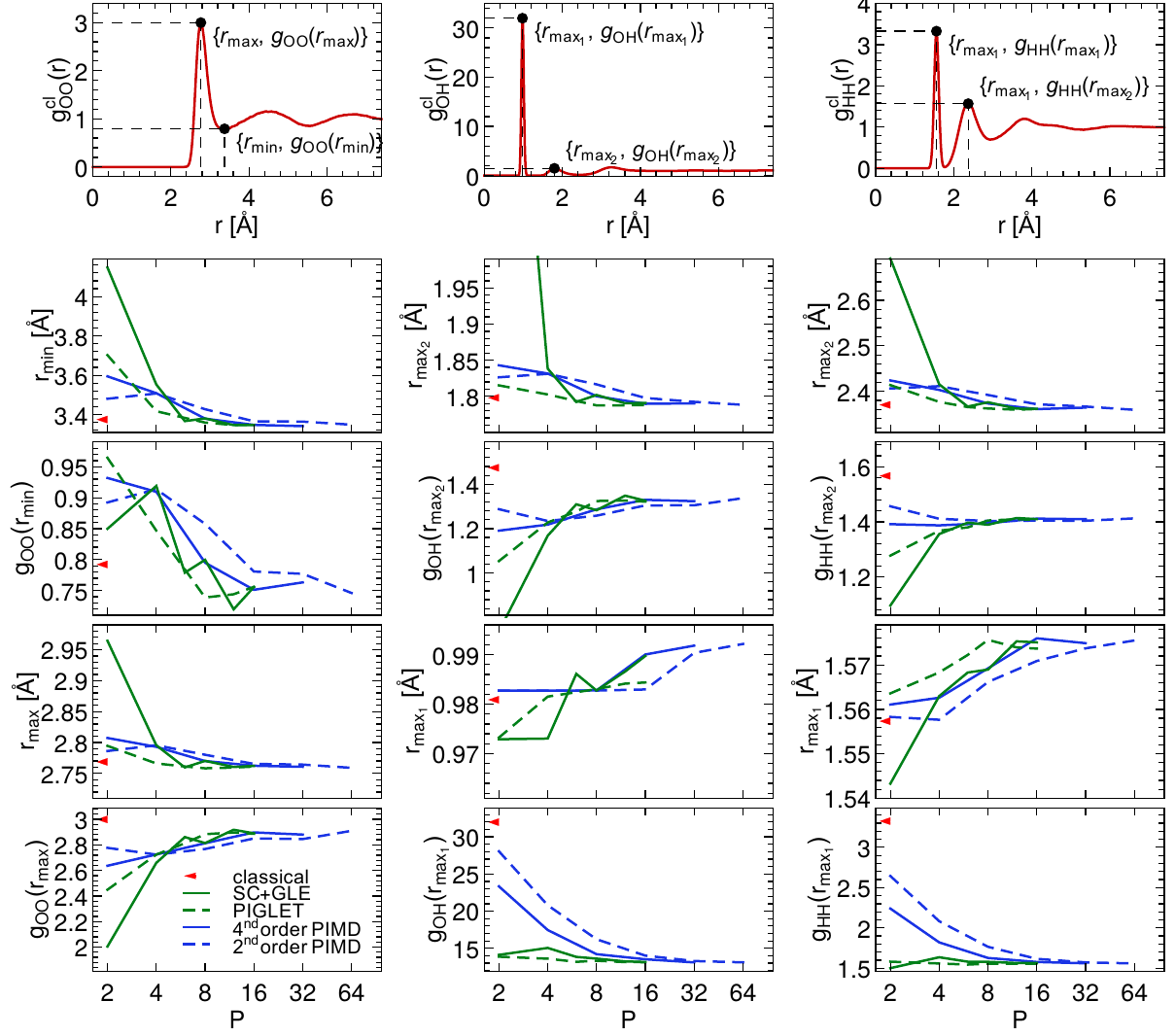}
\caption{\label{fig:gofr} 
The top panels show the radial 
distribution functions for O-O (left), 
O-H (center), and H-H (right) for the classical simulations along 
with some characteristic points whose convergence as a function 
of the number of replicas $P$ is plotted in the four lower panels for each case. The red arrows
show the corresponding values for a purely classical simulation.
}
\end{figure*}

\subsection{Radial distribution functions}
\label{sub:sc-rdf}

The radial pair correlation functions $g(r)$ represent 
the most frequently used indicators of the structure of water. 
Figure~\ref{fig:gofr} shows the convergence of 
a few key features in the O-O, O-H and H-H 
correlation functions. As it has already been 
noted~\cite{gane+13prb}, for Trotter PIMD there is an 
interesting non-monotonic convergence behavior of the $g_\text{OO}$
distribution function, that gets less structured when
going from classical to 2 and 4 beads, and then becomes more
structured when it approaches convergence. Such a trend
can be seen as a manifestation of the competition between
quantum effects in different vibrational modes, that progressively
converge as the number of replicas is increased. 
Overall, the convergence of the radial distribution functions
with $P$ is fully consistent with what is
observed for the energy estimators. The more strongly
quantized degrees of freedom -- such as the O-H stretch -- 
show the slowest convergence, and the most dramatic improvements
with SC path integrals and colored-noise techniques. 
GLE methods give very good agreement with 6-8 beads, but if
a very high accuracy is required -- as it is often needed
in case of radial distribution functions, for which 
changes in the peak shapes of a few percent can be
significant -- SC PIMD with 16 beads gives the best 
performance/cost ratio.

\begin{figure}[hbtp]
\includegraphics[width=1.0\columnwidth]{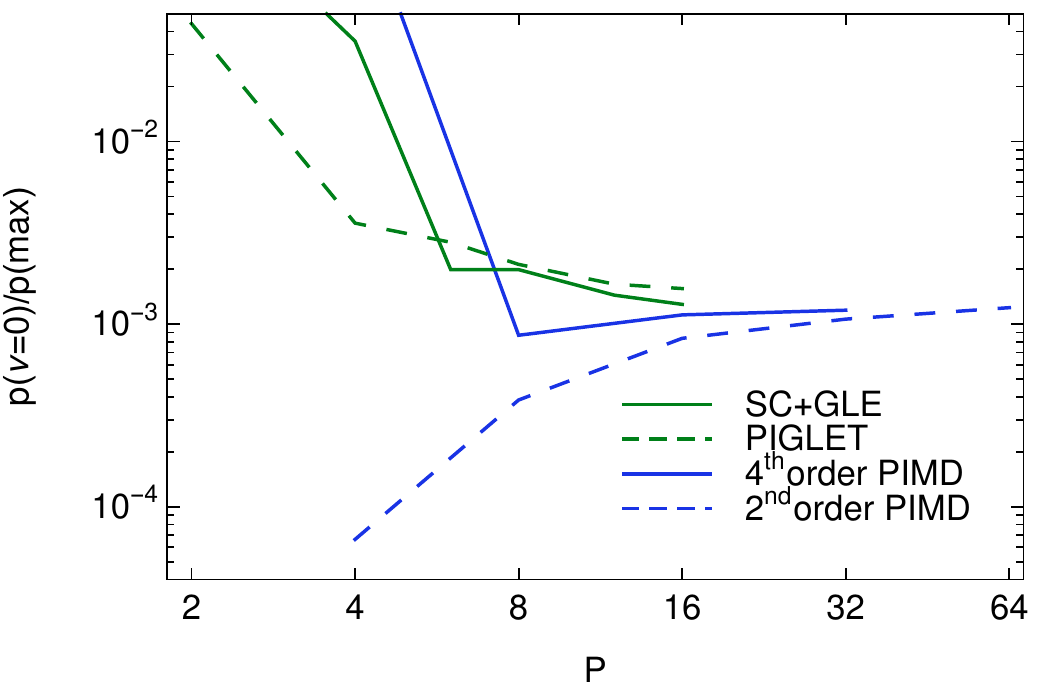}
\caption{\label{fig:nu0}
\mc{
Relative probability for observing proton delocalization
over a H-bond. $p(\nu)$ is the probability density 
relative to the proton-transfer coordinate $\nu$,
and the plot reports the ratio between $p(0)$
and the most likely H-bond configuration $p(\text{max})$, 
as a function of the number of replicas $P$. Note the the 
convergence is slower than for the energy in
Fig.~\ref{fig:energies}.}}
\end{figure}

\subsection{H-bond fluctuations}

One of the most remarkable effects of
quantum fluctuations in room-temperature
water is the occurrence of transient
self-dissociation events, in which a 
quantum fluctuation momentarily brings 
a proton closer to the acceptor oxygen atom
than to the oxygen it is covalently bound to~\cite{ceri+13pnas}. 
The extent of these fluctuations is 
a particularly challenging quantity
to compute, because of the small fraction
of particles that undergo such broad
excursions at any given time, 
the strong anharmonicity of the 
potential in this region, and the dependence on the
level of electronic structure theory~\cite{wang+14jcp}. 
Figure~\ref{fig:nu0} shows the  
probability of having a proton mid-way
between the donor and acceptor oxygen 
relative to the probability of the most
common value of the proton transfer coordinate $\nu$. 
The convergence is 
very slow for all methods, with the
exception of SC PIMD - although \vk{for $P\le 4$}
fourth-order methods give
dramatic over-estimation of these 
fluctuations. Colored-noise methods 
do accelerate convergence, but 
tend to yield too high fluctuations.
For $P=6$, PIMD would underestimate the fluctuations by 
a factor of 5, whereas
PIGLET provides a too high value by a factor of 2.
SC+GLE improves the convergence relative to 
PIGLET -- an advantage that is 
however less significant when one considers the 
increase in computational cost.
Even when predicting strongly anharmonic fluctuations, 
GLE techniques make it possible
to reach semi-quantitative accuracy quickly, and 
fourth-order path integrals are useful to reach 
full convergence in the asymptotic regime.

\begin{figure}[hbtp]
\includegraphics[width=1.0\columnwidth]{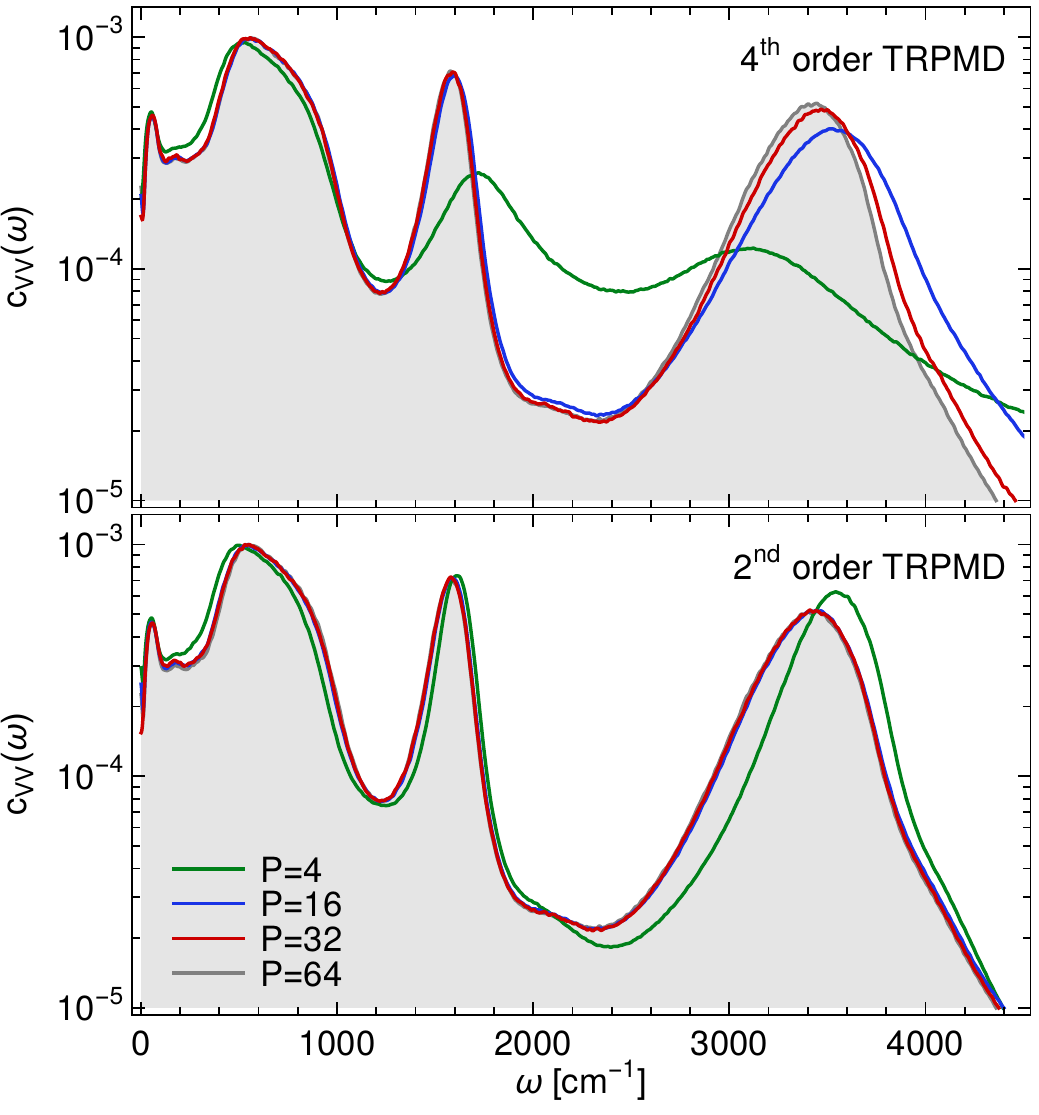}
\caption{\label{fig:cvv} Vibrational densities
of states (Fourier-transforms of the velocity-velocity
correlation functions) for TRPMD simulations of 
liquid water at 300~K, using a second-order Hamiltonian
(lower panel) and a fourth-order Hamiltonian (upper panel).
Simulations with $P=4,16,32$ (green, blue, red) are
compared with a fully-converged Trotter TRPMD simulation
(gray, shaded). 
}
\end{figure}

\subsection{Vibrational density of states}

This far we have focused exclusively 
on static, time-independent properties. 
The path integral formalism is of a statistical
mechanical nature, and strictly speaking no
dynamical observable can be inferred. 
That said, several methods inspired by PIMD
(such as centroid molecular dynamics, 
CMD~\cite{cao-voth93jcp,cao-voth94jcp} and 
ring polymer molecular dynamics, RPMD~\cite{crai-mano04jcp,habe+13arpc})
have been proposed to approximately estimate
diffusion coefficients, vibrational spectra
and other time-dependent quantities. 
For these benchmarks we will focus on thermostatted 
RPMD (TRPMD)~\cite{ross+14jcp}, a simple approach 
that can be seen as combining elements of CMD and RPMD,
alleviating some of their most severe artifacts~\cite{witt+09jcp},
at the price however of a broadening of high-frequency peaks~\cite{ross+14jcp2}.
The idea is just to attach a Langevin thermostat
to ring-polymer modes, with a damping coefficient
adjusted to be proportional to the frequency
of the mode in the free-particle limit. 
For $V=0$ there is no difference between the second
and fourth-order
Hamiltonians, and consequently the TRPMD approach can be
applied in exactly the same way to a fourth-order
simulation.

Generally, one performs (T)RPMD using a number
of replicas that is sufficient to converge 
satisfactorily the static properties to their
quantum values. In Figure~\ref{fig:cvv} we investigate
the convergence of the vibrational density of states
(velocity-velocity correlation spectrum) of water
with increasing numbers of replicas. 
Interestingly, the $c_{vv}(\omega)$ converges
faster than structural properties. 
When using $\mathcal{H}_{P}^{\text{tr}}$,
 $P=16$ is sufficient to obtain
a vibrational spectrum that is indistinguishable 
from the fully-converged limit. On the other hand, 
convergence for $\mathcal{H}_{P}^{\text{sc}}$ 
is dramatically slowed down. This is  
consistent with what observed in 
Ref.~\cite{pere-tuck11jcp} for a harmonic 
potential and the closely-related case of 
Takahashi-Imada path integrals: the physical
vibration is shifted to higher values, and the
discrepancy decays slowly, as $1/P^2$. 
Even at $P=32$ one can observe a significant 
blue shift and broadening of the OH stretch peak
relative to fully converged Trotter TRPMD. 
Although TRPMD based on fourth-order path 
integrals eventually converges to the same 
spectrum as conventional Trotter TRPMD,
it does so very slowly, and so there is no 
advantage in applying fourth-order factorizations
to approximate quantum dynamics. \vk{The SC scheme could however be
used to accelerate the convergence of the mean field centroid
force in the case of fully adiabatic centroid molecular dynamics.\cite{jang-voth01jcp}}

\section{Conclusions}

In the present work we have shown that high-order 
path integral partition functions can be sampled efficiently
by molecular dynamics using a finite-differences 
evaluation of the second derivatives of 
the potential energy. This approach makes
it possible to use these methods \vk{without 
incurring sampling problems} that 
are associated with statistical reweighting.

We benchmarked our method
for the paradigmatic example
of liquid water, using a 
Neural Network potential to model inter-atomic
forces at the DFT level, that
can also describe extreme anharmonicities in the stretch mode of a H-bonded OH. 
We show that our finite-differences 
SC integrator is indeed extremely stable,
and that, even in a room temperature
regime, it enables computational savings 
-- particularly when combined with a multiple
time stepping scheme. 
An explicit MD integrator for high-order
PIMD makes it possible to combine a Suzuki-Chin
factorization with colored noise, and 
approximate quantum dynamics schemes. 
Unfortunately, we find that a combined SC+GLE approach 
does not lead to significant improvements over its 
Trotter (PIGLET) counterpart in this 
temperature regime, due to the 
stronger coupling between different
ring polymer vibrations in the full
SC Hamiltonian, and that
a ``fourth-order RPMD" scheme
has poor performances for dynamical properties. 
The possibility of performing molecular dynamics
with fourth order path integrals, with arbitrary potentials and 
a moderate computational overhead,
provides an additional tool for assessing
accurately the impact of quantum fluctuations
of nuclei in the condensed phase, and might 
open the way to new approaches
to reduce even further the computational 
cost of this kind of simulations at and below
room temperature.

\begin{acknowledgments}
We acknowledge financial support by the Swiss National
Science Foundation (project ID 200021-159896), and computational
time from CSCS under the project IDs s466, s553, and s618.
JB is grateful for funding by the DFG cluster of excellence RESOLV (EXC 1069) 
and for a DFG Heisenberg fellowship (Be3264/6-1).
Discussions with Tobias Morawietz are gratefully acknowledged.
\end{acknowledgments}

\appendix

\section{Symplectic behavior of the
finite-differences SC integrator}
\label{sec:sympl}

One of the crucial properties that make
the scheme we introduce in this work viable is the fact that the velocity-Verlet integrator 
remains symplectic even when using a 
finite-differences scheme to compute
the SC forces. To illustrate this, let us 
consider a simplified one-dimensional
example that captures the essence of 
the method. The propagation over a time
step $\Delta t$ of the SC force alone
corresponds to the steps 
\begin{equation}
\begin{split}
\tilde{p}=&p-\Phi(q)\\
q'=&q+\Delta t \tilde{p}\\
p'=&\tilde{p} - \Phi(q'),\\
\end{split}\label{eq:sc-brief}
\end{equation}
where we introduced the shorthand
\begin{equation}
\Phi(q)=\frac{\Delta t}{2}
\left[c_1 V(q) + c_2 \frac{V(q+\epsilon f(q))-V(q-\epsilon f(q))}{2\epsilon}\right].
\end{equation}
The expression~\eqref{eq:sc-brief} 
\mc{(as well as the corresponding expression
    for the forward FD integrator)} is 
clearly time-reversible due to \mc{the fact
that $\Phi$ only depends on the instantaneous
value of $q$. For the same reason, 
the integrator is symplectic, i.e.
the determinant of the Jacobian of
the propagator  $\left|J\right|=\nicefrac{\partial q'}{\partial q}\, \nicefrac{\partial p'}{\partial p }
-\nicefrac{\partial q'}{\partial p}\,  \nicefrac{\partial p'}{\partial q}$ has 
a unit value.
}
It is straightforward to see that this is the case, by noting that 
\begin{equation}
\begin{split}
\frac{\partial q'}{\partial q}=&
1-\Delta t \Phi'(q) \\
\frac{\partial p'}{\partial p}=&
1-\Delta t\Phi'(q')\\
\frac{\partial q'}{\partial p}=&
\Delta t \\
\frac{\partial p'}{\partial q}=&
-\Phi'(q)-\Phi'(q')\left[1-\Delta t \Phi'(q)\right]\\
\end{split}
\end{equation}

\section{Multiple-time step Suzuki-Chin PIMD}
\label{sec:mts-sc}

Applying a multiple time step (MTS) procedure to a 
SC PIMD simulation is not completely trivial. 
In the presence of a splitting of the physical 
potential $V$ into a short-range (cheap) $V_\text{sr}$
and long-range (expensive) $V_\text{lr}$ potential,
the forces arising from the two terms get non-linearly
coupled. 
Differentiating the 
$\left|\mathbf{f}_\text{sr}+\mathbf{f}_\text{lr}\right|^2$ 
term leads to mixed-derivatives containing
the short-range Hessian projected on the long-range force
and the long-range Hessian projected on the slow force.
In order to obtain significant savings, one 
would have to keep in the outer loop all the terms
that contain long-range forces, including those mixed
with the short-range Hessian. \mc{These terms would 
fluctuate on a fast time scale, thus 
reducing the stability range of the outer time step
and limiting the achievable computational savings.}

\begin{figure}[hbtp]
\includegraphics[width=1.0\columnwidth]{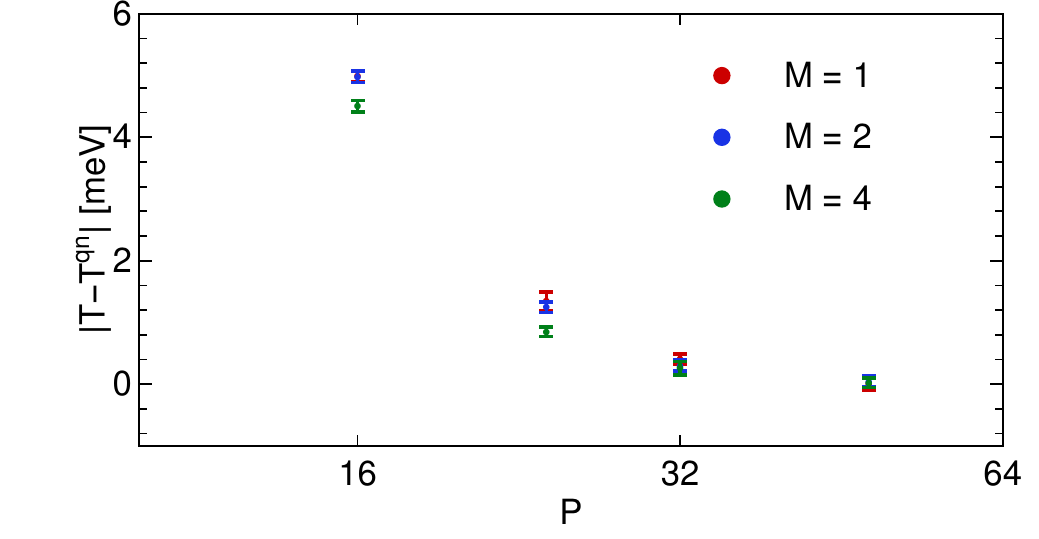}
\caption{\label{fig:mts-sc}
Convergence of the quantum kinetic energy \vk{per molecule} for a 
simulation of liquid water at 300K as a
function of $P$, using a 
finite-difference SC integrator and a MTS 
scheme.
The first derivatives of the physical potential
have been evaluated every $\Delta t=0.25$fs, 
whereas the finite-difference term has
been evaluated once $M$ time steps.  
}
\end{figure}

There is however another aspect for which a MTS procedure 
can be beneficial in the context of SC PIMD. 
The force-dependent term in the SC Hamiltonian is scaled
by a pre-factor that becomes smaller as $P$ is increased. 
One can then apply a MTS splitting in which the (weighted)
Trotter force $w_j{\mathbf{f}}^{(j)}$
is applied in the inner loop, and the
hard-to-compute $w_j d_j\tilde{\mathbf{f}}^{(j)}$ term is applied
in the outer loop, every $M$ steps.
As shown in Figure~\ref{fig:mts-sc}, this
method becomes more and more accurate as
SC-PI approaches full convergence -- which
is the regime in which high-order
path integrals give the greatest advantage.
Up to an outer time step of about 1fs 
(corresponding to an overhead of about 10\%{} 
relative to Trotter PI with the same number
of beads, \mc{ when using a forward-FD scheme}) the
inaccuracy due to the MTS splitting is smaller
than the residual finite-$P$ convergence error. 
Pushing $M$ to 
even higher values leads to more pronounced
errors, without significantly reducing the 
computational cost.

\section{Estimation of equilibrium averages}
\label{sec:estimators}

Let us first consider the thermodynamic expression for
position-dependent operators, that can be derived using the identity
\begin{equation}
    \braket{X} = -\frac{1}{\beta Z^{\text{sc}}} \frac{\partial Z^{\text{sc}}\left(V+\lambda X\right)}{\partial \lambda}|_{\lambda=1}
\end{equation}
which in case of the potential energy gives
\begin{equation}
    \mathcal{V}^\text{TD}_P\left(\mathbf{q}\right)  = \frac{1}{P} \sum_{j=0}^{P-1}  w_j V\left(\mathbf{q}^{(j)}\right) + \frac{2}{P}\sum_{j=0}^{P-1} \sum_{i=0}^{N-1}\frac{w_j d_j}{m_i \omega_P^2}\left|\mathbf{f}_i^{(j)}\right|^2.
\end{equation}
Note that this is very similar, but not identical, to the 
potential term $\mathcal{V}^\text{sc}_P$ entering the
SC Hamiltonian. 
OP-method estimators exploit
the fact that the SC factorization can be seen as an imaginary-time
propagator that goes over two replicas at a time -- so that even
beads sample the proper quantum mechanical distribution of $\mathbf{q}$. As a consequence, position-dependent OP estimators
are very simple, and take the general form
\begin{equation}
     \mathcal{X}^\text{OP}_P(\mathbf{q}) = \frac{2}{P} \sum_{j\in \text{even}} X\left(\mathbf{q}^{(j)}\right).
\end{equation}
For instance, for the potential this yields
\begin{equation}
     \mathcal{V}^\text{OP}_P\left(\mathbf{q}\right) = \frac{2}{P} \sum_{j\in \text{even}} V\left(\mathbf{q}^{(j)}\right).
\end{equation}

\begin{figure}[hbtp]
\includegraphics[width=1.0\columnwidth]{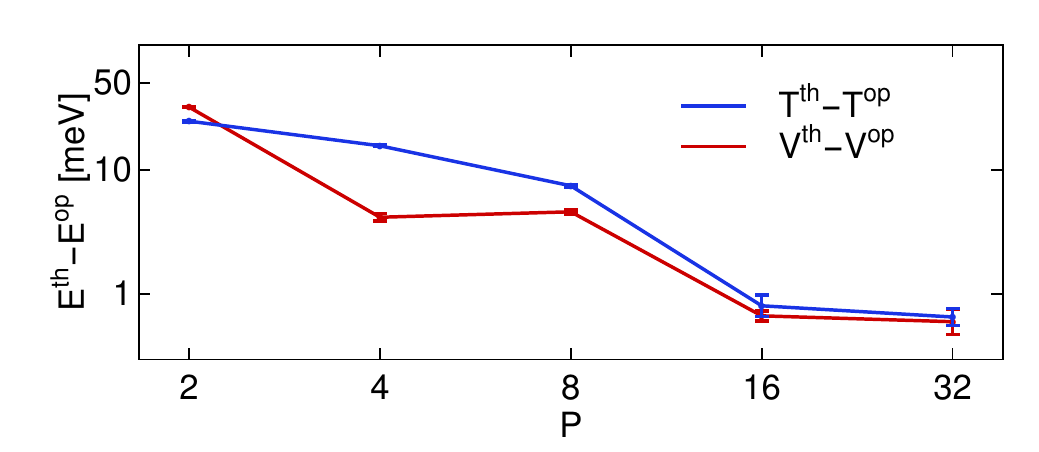}
\caption{\label{fig:tdop}
Difference between the TD and OP fourth-order
estimators of the potential and kinetic
energy per molecule for a simulation of room-temperature
liquid water, as a function of the number
of replicas.
}
\end{figure}

Moving on to the case of the kinetic energy estimators,
for the TD-method the identity we obtain is
\begin{equation}
    \braket{\mathcal{T}} = \frac{1}{\beta Z^{\vk{\textrm{sc}}}} \sum_{i=1}^{N} m_i \frac{\partial Z^{\vk{\textrm{sc}}}}{\partial m_i},
\end{equation}
which can be written as
\begin{equation}
\begin{split}
    \mathcal{T}^\text{TD}_P\left(\mathbf{q}\right)  = &
    \frac{3N}{2\beta} +\frac{1}{2P} \sum_{j=0}^{P-1} \sum_{i=0}^{N-1} \left(\mathbf{q}_{i}^{(j)}-\bar{\mathbf{q}}_i\right)\cdot\mathbf{f}_{i}^{\text{sc},(j)} \\
    &
     + \frac{1}{P}\sum_{j=0}^{P-1} \sum_{i=0}^{N}\frac{w_j d_j}{m_i \omega_P^2}\left|\mathbf{f}_i^{(j)}\right|^2.
\end{split}
\end{equation}
where 
\begin{equation}
    \bar{\mathbf{q}}^{(j)} = \frac{1}{P} \sum_{j=0}^{P-1} \mathbf{q}_{i}^{(j)}
\end{equation}
is the centroid of the ring polymer. The second-derivative
component within the SC force $\mathbf{f}_{i}^{\text{sc},(j)}$ can be computed using the same finite-difference expression~\eqref{eq:finitediff}
that is used to propagate the dynamics. 
The OP estimator for the kinetic energy is effectively equivalent
to the usual centroid virial kinetic energy estimator, evaluated
on just the even beads:
\begin{equation}
     \mathcal{T}^\text{OP}_P\left(\mathbf{q}\right) =
    \frac{3N}{2\beta} +\frac{1}{P} \sum_{j\in\text{even}} 
    \sum_{i=0}^{N-1} 
    \left({\mathbf{q}}_{i}^{(j)}-\bar{\mathbf{q}}^{(j)}\right)\cdot\\{\mathbf{f}}_{i}^{(j)},
\end{equation}
Note that in general the possibility of computing OP 
estimators, that do not contain $\mathbf{f}_{i}^{\text{sc},(j)}$,
is an important feature of SC path integrals, as that avoids
computing second derivatives of the potential in re-weighted
schemes.
As we have seen, the finite-difference approach we use here
makes it trivial to compute the TD estimators, so one can 
verify which flavor converges more rapidly to the 
quantum limit and/or cross-validate simulation results. In the case of 
room-temperature liquid water, we
find that there is little difference
between the convergence of TD and OP,
and that the two estimators become
indistinguishable by the time full convergence
is achieved (see Figure~\ref{fig:tdop}).

\section{Effective temperature curves
for SC+GLE}\label{sec:scgle}

In the harmonic limit, for a physical potential of 
frequency $\omega$, the frequencies
$\omega_k(\omega)$, and the 
eigenvectors ${\mathbf{u}}^{(k)}(\omega)$ of 
the SC Hamiltonian can be obtained by
diagonalizing the dynamical matrix 
$D_{jj'}$, that reads
\[  D_{jj'} =\left\{
\begin{array}{ll}
      2\omega_P^2 +  \omega^2 w_j\left( 1+ 2d_j\left(\frac{\omega}{\omega_P}\right)^2 \right)  & j = j' \\
      -\omega_P^2 & j = j' \pm 1  \\
      0 & \text{otherwise}  \\
\end{array} 
\right. \]
where the cyclic boundary conditions 
$j+P\equiv j$ are implied. The dynamical
matrix, and as a consequence the derivations that follow,
depend on the choice of the parameter $\alpha$
(see Eq.~\eqref{eq:sc-wd}). We will continue from here on assuming $\alpha=0$, but the 
generalization to an arbitrary $\alpha$ is straightforward.

A further complication stems from the
fact that the estimator for the 
fluctuations of $q$ is not just an
average over the coordinates of 
all beads. In fact, one has 
to choose whether to design the GLE
so as to speed up the convergence of 
either the TD or the OP
estimator for $\left<q^2\right>$. Given its simplicity, and the direct connection
with all structural observables, we
opted for the latter choice, that gives
\begin{equation}
\left<q^2\right>=\frac{2}{P}
\sum_{j\in \text{even}} \left<\left[q^{(j)}\right]^2\right>
=\frac{1}{P}\sum_k
U_k \left<\left[\tilde{q}^{(k)}\right]^2\right>
\end{equation}
where 
\begin{equation}
U_k=2\sum_{j\in \text{even}}\left|u_j^{(k)}\right|^2
\end{equation}
gives the weight of the $k$-th mode 
on the displacement of the even beads.

The design of the effective-temperature curve $T^\star(\omega)$
can then proceed in a similar way to what was done in Refs.~\cite{ceri+11jcp,ceri-mano12prl}:
considering that for a classical oscillator $\left<q^2\right>=\nicefrac{1}{m\beta\omega^2}$
one can write the functional equation
\begin{equation}
\frac{\hbar}{2\omega k_B}\coth \frac{\hbar\omega\beta}{2} =
\frac{1}{P}\sum_k U_k(\omega) \frac{T^\star(\omega_k(\omega))}{\omega_k(\omega)^2}.
\label{eq:tstar-sc}
\end{equation}
Here we made explicit the dependence of the eigenvector coefficients and of the normal
modes frequencies on the physical frequency of the underlying potential. 
Solution of Eq.~\eqref{eq:tstar-sc} can be obtained by singling out 
the lowest-lying \vk{NM}, obtaining the iteration
\begin{equation}
\begin{split}
T^\star(\omega_0)  =& \frac{\omega_0^2}{U_0} \left[\frac{P\hbar}{2\omega(\omega_0) k_B}\coth \frac{\hbar\omega(\omega_0)\beta}{2} - \right.\\
& \left.
\sum_{k>0} U_k(\omega_0) \frac{T^\star(\omega_k(\omega_0))}{\omega_k(\omega_0)^2}\right]
\end{split}
\label{eq:tstar-iter}
\end{equation}
that can be made to converge with an appropriate mixing
scheme~\cite{ceri+11jcp} and with the starting condition
\begin{equation}
T^\star(\omega_0)= \frac{\hbar \omega_0}{\sqrt{6} k_B} \coth \left(\frac{\beta \hbar \omega_0}{\sqrt{6} P}\right)
\end{equation}
Yet another complication associated with using a GLE in connection
with SC path integrals is that the lowest normal-mode frequency is not
equal to the physical frequency, as in Trotter PIMD. For this 
reason, one needs to invert the $\omega_0(\omega)$ relation to
find what is the physical frequency that corresponds to the argument
of $T^\star$ we are solving for. In a similar way, one can then 
obtain the frequencies of the higher \vk{NMs} as a function of 
the lower frequency $\omega_0$, which eventually makes it possible
to solve numerically the iteration in Eq.~\ref{eq:tstar-iter}. 
In fact, it is possible to give a closed (albeit cumbersome)
expression for such inverse relation~\cite{brainthesis}
\begin{equation}
\begin{split}
\omega(\omega_0)=&\frac{2^{2/3} A^{2/3}+\sqrt[3]{A} \left(4 x_0^2-6 P^2\right)+4 \sqrt[3]{2} x_0^2 \left(3 P^2+2 x_0^2\right)}{8\beta\hbar \sqrt[3]{A}}\\
A=&27 P^6-72 P^2 x_0^4+16 x_0^6 +3 \left[81 P^{12}-432 P^8 x_0^4+\right.\\
&+ \left.384 P^4 x_0^8-384 P^2 x_0^{10}\right]^{1/2}\\
x_0=&\beta\hbar\omega_0.
\end{split}
\end{equation}

\begin{figure*}[hbtp]
\includegraphics[width=1.0\textwidth]{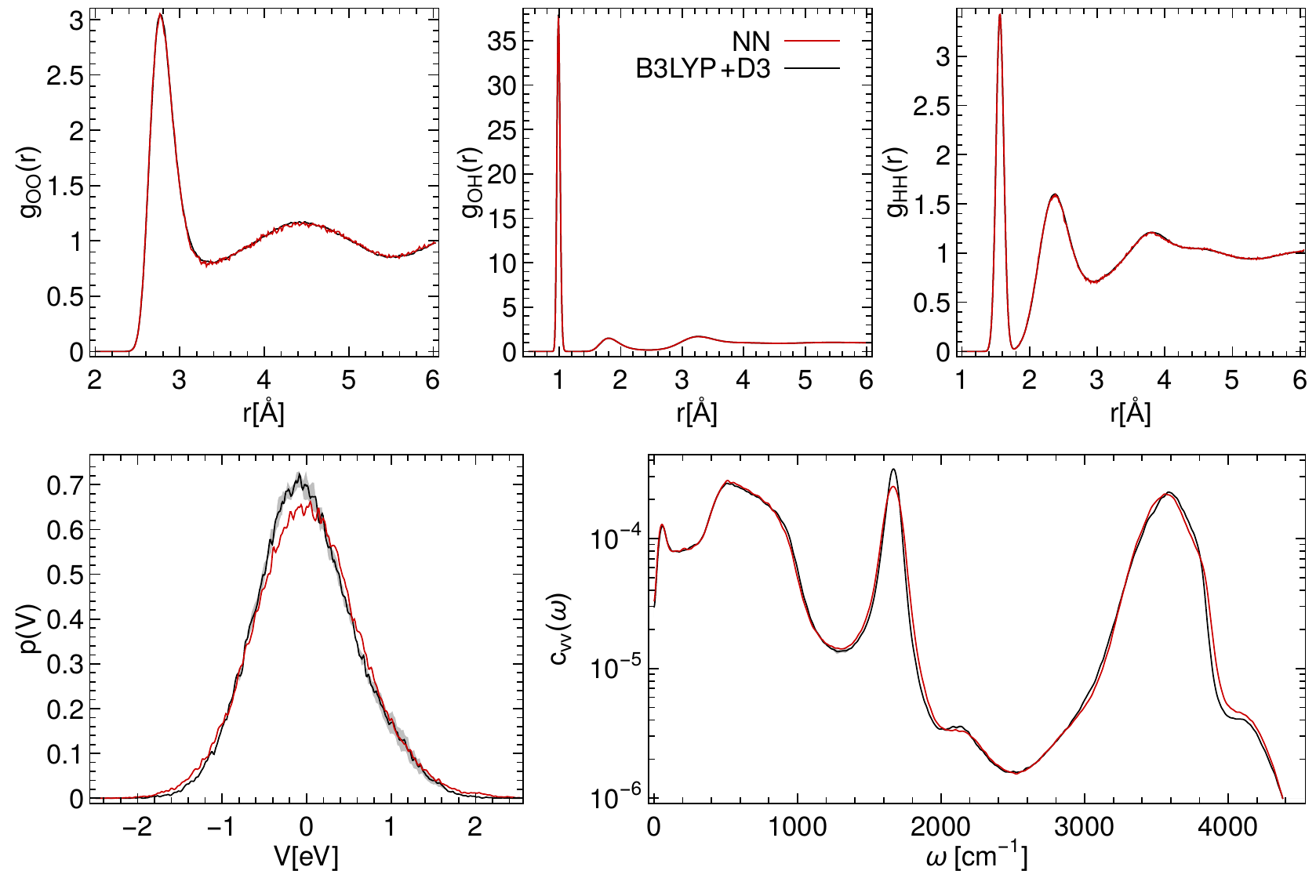}
\caption{\label{fig:b3lyp-nn}
Comparison between structural and dynamical properties
from a \emph{NVT} simulation of 64 water molecules
at 300K using B3LYP+D3 (black) and the NN fit (red). 
The top panels show the pair correlation functions, 
the bottom-left panel shows the distribution of potential
energy fluctuations around the average, and the 
bottom-right panel shows the vibrational density of states
(velocity-velocity correlation function)
}
\end{figure*}

\section{Accuracy of the neural-network fit}
\label{sec:nn-test}

The high-dimensional neural-network potential we use in the present work employs the framework developed by Behler and Parrinello~\cite{behl-parr07prl,behler2011,behler2014}. Our fit, which follows the NN water potential reported by Morawietz et al. for the case of generalized gradient approximation energetics~\cite{morawietz2016}, has 
already been shown to provide excellent agreement with 
both the underlying B3LYP+D3 data it has been fitted to
and with experimental quantities, when computing 
quantum mechanical observables such as the nuclear
kinetic energy and isotope fractionation 
ratios~\cite{chen+16jpcl}. Here we will just present 
a few additional diagnostics to demonstrate that
it is similarly accurate when it comes to structural
and dynamical observables. We used classical MD
simulations, for which we could accumulate more than 
60~ps of ab initio trajectory (4 independent runs of
15~ps each) with 64 water molecules at
constant temperature and volume conditions, so we 
can perform a statistically meaningful comparison.
We applied a weak global thermostat~\cite{buss+07jcp}, 
so as to also be able to extract information on 
dynamical properties.

As it can be seen in Figure~\ref{fig:b3lyp-nn},
radial distribution functions (RDF) obtained  from
NN runs are completely indistinguishable
from the reference ab initio simulation.
It ought to be noted that although 
pair correlation functions are often used
to demonstrate the reliability of an
approximate simulation protocol, they 
constitute a very integrated measure,
and success in reproducing RDFs is 
a necessary but not sufficient criterion 
to establish the equivalence of two
inter-atomic potentials. 
The lower panels in Figure~\ref{fig:b3lyp-nn} show
two more challenging tests: the histogram
of potential energy (that also directly
relates to the - classical - heat capacity)
and the density of states, as given by 
the velocity-velocity correlation spectrum. 
Although in these cases one can appreciate
differences between the ab initio reference
and the NN fit (with the latter showing
slightly larger energy fluctuations, and
a noticeably broader bend peak), the 
agreement is excellent. Together with the
accurate reproduction of quantum 
kinetic energy and isotope fractionation 
ratios, these results give us great 
confidence in the quality of the NN 
potential. 
The possibility of studying larger 
boxes for longer simulation times
outweighs by far the minute discrepancies
that are seen for the most stringent
tests of Fig.~\ref{fig:b3lyp-nn}.

\bibliographystyle{aipnum4-1}
\end{document}